\documentclass[amsmath,amssymb,prl,floatfix,reprint,superscriptaddress]{revtex4-2}

\usepackage{graphicx}
\usepackage{dcolumn}
\usepackage{bm}
\usepackage{hyperref}
\usepackage{siunitx}
\usepackage{soul} 

\graphicspath{{}, {/}}

\begin{document} 

\title{Integrated Sideband-Resolved SERS with a Dimer on a Nanobeam Hybrid}

\author{Ilan Shlesinger}
\thanks{Equal contribution}
\affiliation{Department of Physics of Information in Matter and Center for Nanophotonics, NWO-I Institute AMOLF, Science Park 104, NL1098XG Amsterdam, The Netherlands}
\author{Isabelle M. Palstra}
\thanks{Equal contribution}
\affiliation{Institute of Physics, University of Amsterdam, 1098 XH Amsterdam, The Netherlands}
\affiliation{Department of Physics of Information in Matter and Center for Nanophotonics, NWO-I Institute AMOLF, Science Park 104, NL1098XG Amsterdam, The Netherlands}
\author{A. Femius Koenderink}
\email{f.koenderink@amolf.nl}
\affiliation{Department of Physics of Information in Matter and Center for Nanophotonics, NWO-I Institute AMOLF, Science Park 104, NL1098XG Amsterdam, The Netherlands}
\affiliation{Institute of Physics, University of Amsterdam, 1098 XH Amsterdam, The Netherlands}
 
\date{\today}

\begin{abstract}
In analogy to cavity optomechanics, enhancing specific sidebands of a Raman process with narrowband optical resonators would allow for parametric amplification, entanglement of light and molecular vibrations, and reduced transduction noise. We report on the demonstration of waveguide-addressable sideband-resolved surface-enhanced Raman scattering (SERS). We realized a hybrid plasmonic-photonic resonator consisting of a 1D photonic crystal cavity decorated with a sub-20 nm gap dimer nanoantenna. Hybrid resonances in the near-IR provide designer Q-factors of 1000, and $Q/V=(\lambda^3/10^6)^{-1}$, with SERS signal strength on par with levels found in state-of-the-art purely plasmonic systems. 
We evidence Fano-lineshapes in the SERS enhancement of organic molecules, and quantitatively separate out the pump enhancement and optical reservoir contributions.
\end{abstract}


\maketitle







Molecules can exchange energy with incident light through the vibrations of their atomic bonds~\cite{Long2002}, giving rise to inelastic light scattering, called Raman scattering. Surface-enhanced Raman spectroscopy (SERS) uses plasmonic nanoparticles featuring intense field hotspots to enhance the intrinsically weak Raman scattering of molecules by orders of magnitude, and is among the most popular single-molecule sensitive spectroscopy tools~\cite{Fleischmann1974,Jeanmaire1971,Albrecht1977,LeRu2012}. Recently, theoretical efforts have sought to exploit formal analogies between SERS and cavity optomechanics~\cite{Roelli2016MolecularScattering,Schmidt2016QuantumCavities,Benz2016Single-moleculepicocavities,Lombardi2018,Schmidt2017LinkingSERS}, which is the field of precisely controlling and sensing the  mechanical motion of nano and microresonators with light. The molecular optomechanics viewpoint promises to bring the exciting effects that were obtained in nano-engineered MHz/GHz acoustic resonators, such as strong optomechanical coupling, production of non-classical mechanical states, parametric amplification, and  coherent wavelength conversion~\cite{Aspelmeyer2014CavityOptomechanics,Bowen2015}, to THz and mid-IR molecular vibrations. The two latter have recently been demonstrated with molecules~\cite{Lombardi2018,Chen2021,Xomalis2021}, and collective effects in Raman scattering have also been observed~\cite{zhang_2020,Vento2022}.

A central tenet of cavity optomechanics is that so-called sideband-resolved operation is crucial, which is attained when the optical cavity has a linewidth smaller than the mechanical resonance frequency \cite{Aspelmeyer2014CavityOptomechanics,Bowen2015}. It eliminates detrimental backaction, thus allowing ground-state cooling, pure light-mechanical entanglement, and frequency conversion without added noise. 
High-quality factor (Q) dielectric cavities can provide the narrowband spectral structure required for sideband-resolved SERS, but are limited to low electric field confinement, resulting in small optomechanical coupling strengths and very low Raman enhancement compared to plasmonic systems. Conversely, the optical linewidth (Q=10-30) for typical plasmon antennas~\cite{agio2013optical}) exceeds the vibrational frequencies, so that established SERS geometries are not sideband-resolved. A major challenge lies in reaching sideband resolution while maintaining the uniquely strong field enhancement of metal junctions. Furthermore, a significant open question for both traditional SERS and prospective molecular sideband-resolved optomechanics is how to achieve waveguide-addressability.  All SERS demonstrations with few molecules have been limited to free-space configurations, since integrated geometries are rapidly confronted by high background noise emanating from the guiding material~\cite{Peyskens2016SurfacePlatform,Peyskens2018}.

\begin{figure}
  \includegraphics[width=0.9\columnwidth]{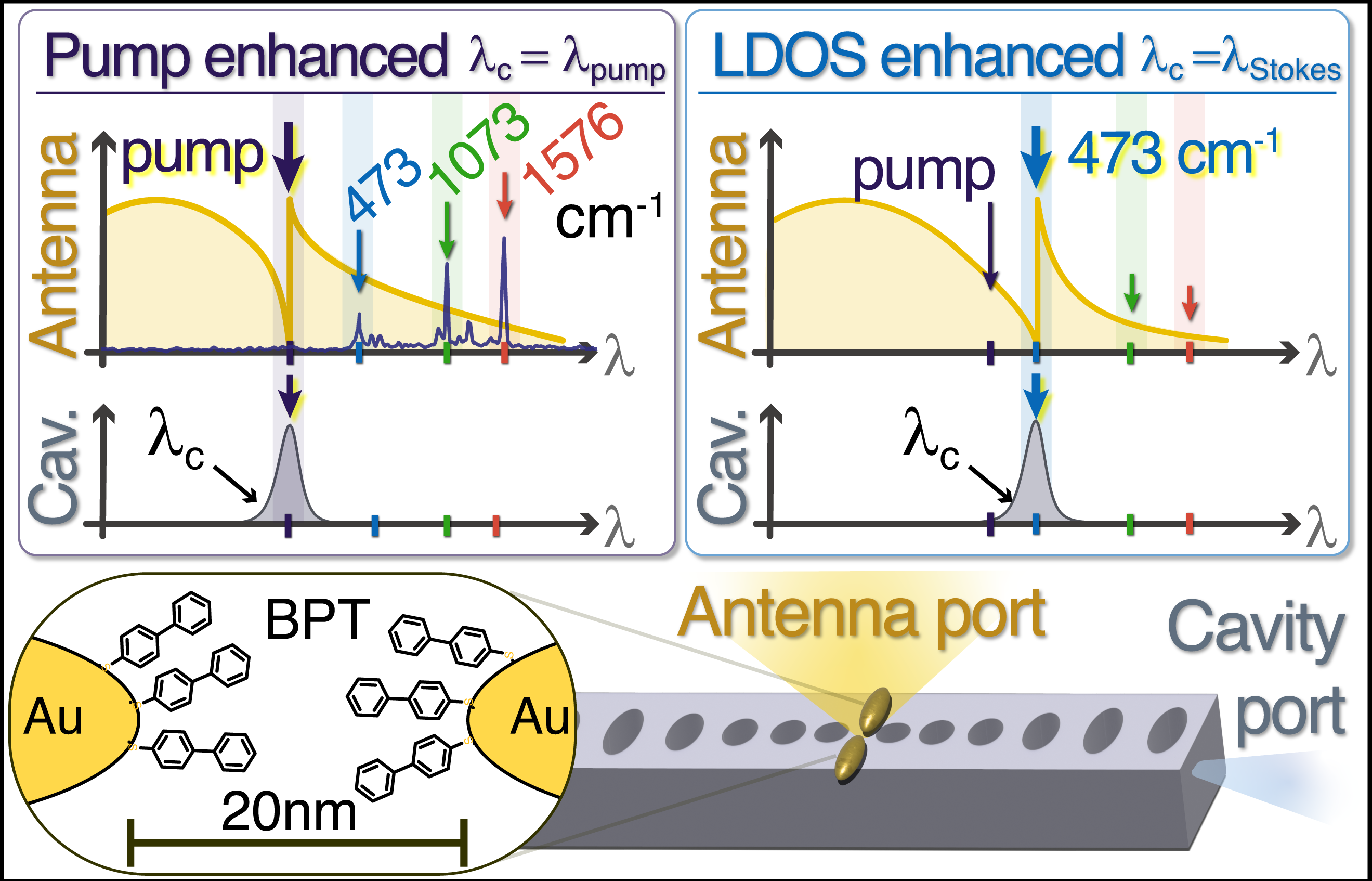}
  \caption{ A gold dimer antenna, decorated with BPT molecules, hybridizing with a Si$_3$N$_4$ photonic crystal nanobeam. Depending on the cavity resonance wavelength, either the pump (left) or a single BPT Raman peak (right) is enhanced by the hybrid resonance. The enhancement lineshape will depend on the addressed optical port, featuring a Fano lineshape for the antenna or a Lorentzian-like shape for the cavity.
  }
  \label{fig:sketch}
\end{figure}

In this Letter, we report on the realization of a proposed new generation of hybrid light resonators~\cite{Xiao2012,Doeleman2016Antenna-CavityLinewidth,Heylman2016} that combine a plasmonic antenna interacting with a dielectric cavity and features narrow and intense resonances~\cite{KamandarDezfouli2017QuantumResonators,Dezfouli2019MolecularModes,Palstra2019HybridCryostat}. We show how it allows for few-molecules sideband-resolved SERS with selective enhancement of single Raman lines and with waveguide excitation and collection capabilities (Fig.~\ref{fig:sketch}). The system combines photonic crystal nanobeam cavities~\cite{Deotare2009} with single gold dimer antennas~\cite{Schuck2005}, aligned within a few nanometers by two-step lithography. 
The hybridization of a broad plasmonic resonance with the sharp cavity peak results in sharp Fano features in the optical reservoir~\cite{KamandarDezfouli2017ModalSystem,Medina2021}, i.e., in the local density of optical states (LDOS) landscape felt by the Raman species with which we functionalized the gold antennas.  We present a study using a confocal Raman spectroscopy setup (see Supplementary Material (SM)~\footnote{See Supplemental Material at [URL will be inserted by publisher], for fabrication and experimental details, mode simulations versus geometry,  and semi-analytical lineshape modelling, which includes References 31-37})\nocite{Yan2018,Vericat2014,Ahmed2021,Xomalis2021,Shlesinger2021,Peyskens2018,Doeleman2016Antenna-CavityLinewidth,Daveau2017,Bowen2015,Roelli2016MolecularScattering,Schmidt2016QuantumCavities,Zhang2020,LeRu2009PrinciplesSpectroscopy,Lombardi2018} with a tunable laser, disentangling the role of pump enhancements and the LDOS landscape, and evidencing waveguide-addressed sideband-resolved SERS.

%

\begin{figure}
  \includegraphics[width=0.88\columnwidth]{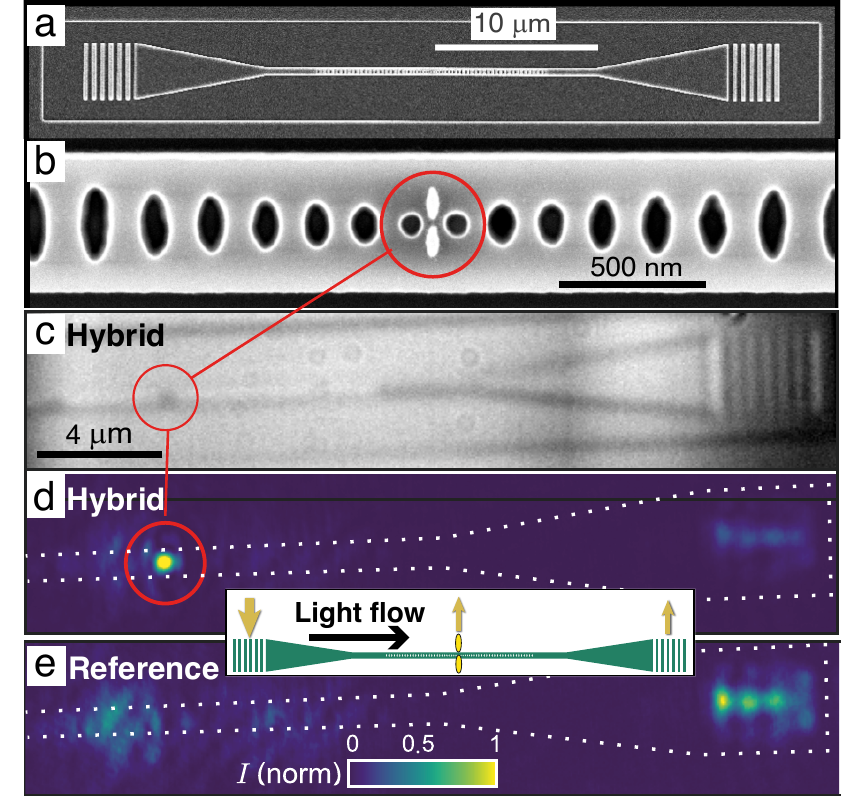}
  \caption{(a) SEM micrograph of a nanobeam cavity with grating couplers on each side. (b) Aligned dimer antenna placed on top and in the center of the cavity. (c) Bright-field White-Light (WL) microscopy image showing the cavity (left-hand part of figure) and grating outcoupler (right-hand). The antenna appears as a black dot. (d) Laser scattering  of the same device and  (e) a device with no antenna. The system is driven by a laser resonant with the cavity, and focused on the left incoupling grating (not visible here). Bright scattering by the antenna demonstrates antenna-cavity coupling. 
  }
  \label{fig:Fabrication}
\end{figure}

SEM micrographs in Fig.~\ref{fig:Fabrication}(a,b) show a representative example of a hybrid composed of a gold dimer antenna with a 20 nm gap placed on top and in the center of a Si$_3$N$_4$ photonic crystal lying on a SiO$_2$ layer (see SM for design and fabrication details~\cite{Note1}).
Two gratings on each side of the nanobeam allow coupling light from free-space to the waveguide. 
While the aligned multi-step electron-beam lithography required to realize these samples is challenging, the robustness of our fabrication protocol is evident from the fact that our study is based on optical measurements on over 280 of such hybrid resonators. 
Design quality factors predict $Q_c\simeq2500$ and mode volumes $V_c\sim 3\lambda^3$ for the nanobeam and $V_a=10^{-4}\lambda^3$ for the dimers, resulting in $Q\sim10^3$ and $V\sim2\times 10^{-3}\lambda^3$ for hybrids~\cite{Doeleman2016Antenna-CavityLinewidth}.  The resulting Q/V ratio of $10^{6}$ in units of $\lambda^3$ is on par with state-of-the-art plasmonic resonators~\cite{Baumberg2019} and photonic crystal cavities~\cite{Deotare2009}, but with the unique asset that Q can be chosen at will from $10^2$ to $10^3$.

\begin{figure}
  \includegraphics[width=0.9\columnwidth]{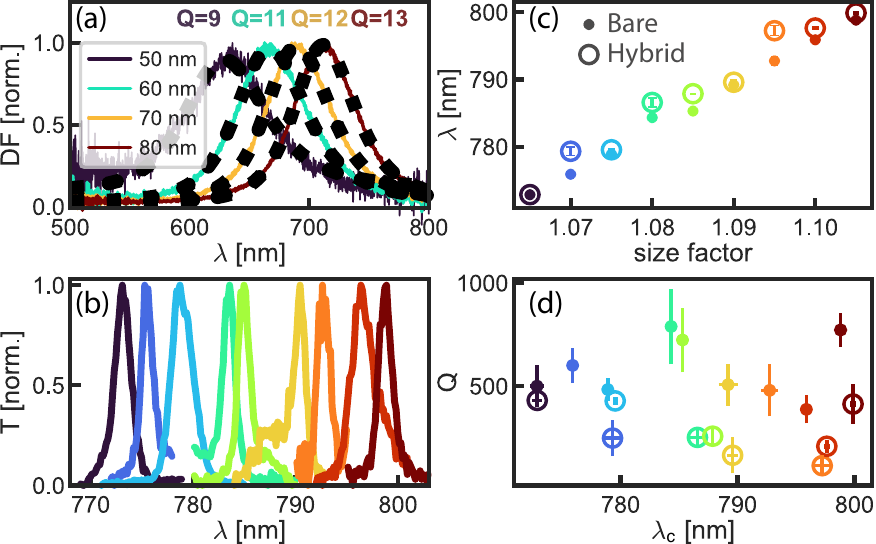}
  \caption{(a) DF spectra  of bare gold dimers. Resonances redshift with antenna size. Q's are extracted from squared Lorentzian fits (dotted lines). (b) Transmission of the bare nanobeam cavities, 9 different designs were made with 4 devices studied per design. Increasing the in-plane photonic crystal hole radius and period (scaling parameter on $x$-axis) tunes  the cavity resonance (c). In the presence of the antenna, the hybridized cavity mode is red-shifted and broadened (open circular markers) (c-d). Errorbars are standard deviations from four different but nominally identical beams/hybrids.
  }
  \label{fig:DFandWL}
\end{figure} 
The first experimental signatures of hybridization are readily observed in the spatial scattering of the cavity presented in Fig.~\ref{fig:Fabrication}(c-d).
In panels (d, e), a tunable diode laser, resonant with the cavity (ca. \SI{780}{\nano\meter}), is aimed at the left incoupling grating, just outside the field of view. 
For the hybrid system, there is bright scattering from the antenna (bright dot in (d); the antenna appears as a dark spot in bright-field microscopy (c)). On the contrary, in absence of an antenna, the light is mainly outcoupled from the second grating (Fig.~\ref{fig:Fabrication}(e)).
This shows that it is possible to drive the gold dimer antenna through the nanobeam structure, implying optical coupling between the cavity and the plasmon resonances. This coupling provides frequency structure in hybrid LDOS enhancement and waveguide addressability of the plasmonic hot spot.

Fig.~\ref{fig:DFandWL}(a,b) reports on the measured optical resonance properties of the individual constituents, that is, antenna and cavity, and of the hybrids. Bare antenna darkfield (DF) spectra confirm that the dimer antennas feature a dipolar resonance with $Q\sim10$ and polarized along their long-axis. 
Fig.~\ref{fig:DFandWL}(b) shows transmission spectra of antenna-free nanobeam cavities. The resonance wavelength can be tuned from 772 to 800 nm by changing the position and width of the holes by 4\% size difference (Fig.~\ref{fig:DFandWL}(c), full dots). 
As shown in Fig.~\ref{fig:DFandWL}(d) in closed symbols, the bare cavity modes have measured $Q\sim600$ with strong variability due to fabrication imperfections, and with the highest values obtained equal to $Q\simeq1000$. 
The parameters of the hybridized cavity with the antennas are shown with circled markers. Although the spread in the nanobeam cavity resonance wavelength $\lambda_c$ and Q due to fabrication fluctuations makes it difficult to directly measure cavity perturbation effects by comparing different devices, we find that hybridized resonances appear systematically redshifted and broadened compared to bare cavities, as expected from cavity perturbation theory~\cite{Schwinger1943}, with a reduction in Q towards 200-300 range. These Q, corresponding to ca. 65 cm$^{-1}$ linewidths, are ideally suited for sideband-resolved SERS
: sufficiently narrow to select single low frequency vibrational lines, yet wide enough to encompass typical Raman linewidths in full~\cite{Long2002}. 
 
\begin{figure}
  \includegraphics[width=\columnwidth]{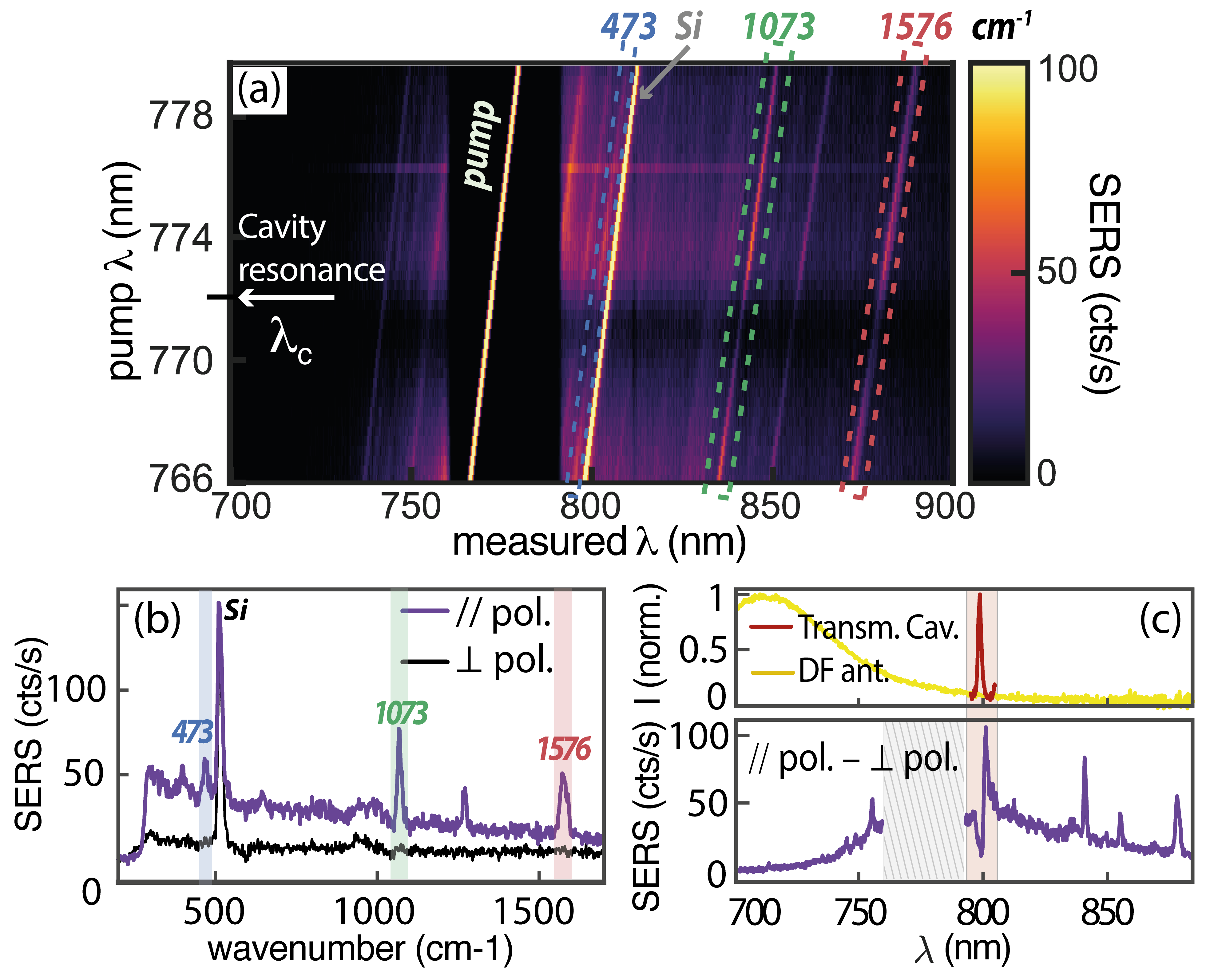}
  \caption{ (a) SERS on a hybrid for pump wavelengths scanned from 766 to 780 nm.  The Fano response of the hybridized antenna can be seen as a lowering and raising of SERS intensity when passing through the cavity resonance frequency (pump enhancement effect, near 772 nm). (b) Cross-cut of (a) for $\lambda_\mathrm{pump}=774\,\mathrm{nm}$ (purple line). Raman scattering from the BPT is only detected when the laser is polarized parallel to the dimer. For orthogonal polarization, only the Si signal from the substrate is seen (black curve).
  (c) DF and transmission spectrum of a hybrid where the cavity resonance at 800 nm lies in the Stokes sideband (LDOS enhancement). Bottom: SERS spectrum featuring a clear Fano resonance  at the hybrid resonance (silicon background subtracted). 
  }
  \label{fig:Raman1st}
\end{figure}

To this end, we functionalize the antennas with biphenyl-4-thiol (BPT) molecules which form self-assembled monolayers on gold, and exhibit distinct Raman scattering~\cite{Note1,Bain1989}. 
Fig.~\ref{fig:Raman1st}(a) shows a set of Raman spectra recorded on a hybrid as function of both detection wavelength and pump wavelength. Here, excitation and collection are done through the antenna port and the laser is polarized parallel to the dimer long axis.
As the pump laser is swept from 766 to 780 nm (residual transmitted pump light through the notch filter visible), the BPT Raman lines (lines of interest highlighted in dashed boxes)  maintain a constant shift from the laser frequency. In Fig.~\ref{fig:Raman1st}(a) the cavity resonance lies at 772 nm. While scanning the laser through that wavelength, the hybrid Fano response of the antenna at the cavity frequency results in a Fano shaped pump enhancement, reducing the pump field enhancement at the Fano dip around 771 nm and increasing it at the Fano peak around 774 nm. The intensity of the electronic Raman scattering ~\cite{Mertens2017} (broad Raman background) and the narrow BPT Raman peaks show this trend in the pump enhancement, most noticeably as the dark region around 771 nm.  Instead, the silicon peak (bright line at 520~cm$^{-1}$) emitted by the underlying substrate has a constant intensity, evidencing that the signal modulation is actually due to the hybrid resonance and not to an overall change in pump laser intensity.
To confirm that the SERS signal comes strictly from BPT molecules at the antenna, we plot in Fig.~\ref{fig:Raman1st}(b) a cross-cut of the Raman spectrum recorded at $\lambda_\mathrm{laser}=774\,\mathrm{nm}$ (purple line) and a reference case with pump polarization orthogonal to the antenna long axis (black line). SERS signal is observed only for the parallel case. In the orthogonal configuration no BPT Raman peaks appear and only the Raman line from the silicon substrate is observed. The intermediate glass and PMMA cladding do not contribute to the Raman signal.  Considering that the BPT counts in the reference case are below noise levels of 1 count/s, we deduce a lower bound on Raman enhancement of $1.1\times10^4$, limited by our 20 min. integration time. The actual enhancement is estimated to be 2-3 orders of magnitude larger using a semi-analytical model presented in the SM~\cite{Note1}.
 
To highlight the effect of hybrid LDOS, Fig.~\ref{fig:Raman1st}(c) shows a Raman spectrum from a different hybrid with $\lambda_c=800$ nm that now lies in the Stokes sideband. The top diagram shows the relevant antenna DF spectrum and cavity transmission. Interference of the narrow-cavity resonance and the broad dimer response results in a structured LDOS with a Fano lineshape. This Fano resonance is imprinted on the broad electronic Raman scattering background of the gold, around 800 nm wavelength (bottom panel), and the Fano peak is tuned to enhance the BPT line at $473\,\mathrm{cm}^{-1}$. Combined, Figs.~\ref{fig:Raman1st}(a-c) qualitatively show that it is possible to superimpose the hybrid resonance selectively with either the pump wavelength or with specific Raman lines. 

\begin{figure}
  \centering
  \includegraphics[width=0.95\columnwidth]{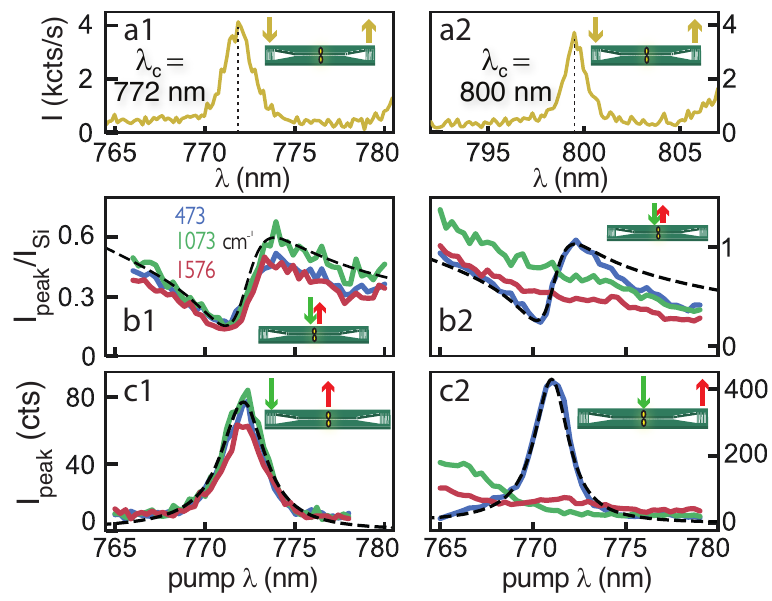}
  \caption{
  Pump (left column) or LDOS (right column) hybrid enhancement. The pump (left) or the 473~cm$^{-1}$ peak (right) are respectively swept through the hybrid cavity resonance.
  (a) WL transmission spectrum of the hybridized cavity resonance. (b,c) Integrated counts of 3 selected BPT Raman peaks intensities for laser wavelengths from 765 to 780 nm. In
  (b) pump and collection is through the antenna and normalized by the Si line, taken as reference.
  (c1) coupling through one grating and collecting at the antenna and (c2) is the inversed configuration. Dashed curves are from a semi-analytical theoretical model~\cite{Note1}.
  }
  \label{fig:Raman2}
\end{figure}

In Fig.~\ref{fig:Raman2}, we quantitatively explore the SERS enhancement engineering achieved with the hybrid resonance, allowing integrated sideband-resolved SERS.
%
First, we discuss the SERS response of hybrids resonant with the pump (panels a1--c1) with a hybrid resonance at $\lambda_c=\SI{772}{\nano\meter}$ (same structure as in Fig.~\ref{fig:Raman1st}(a,b)).
Panels b1 and c1 are excitation spectra reporting integrated counts of the Raman peaks at 473, 1073, and 1576 cm$^{-1}$ (blue, green and red curves) as function of pump wavelength (swept from 765-780 nm at \SI{400}{\micro\watt} power, 20 s integration time per data point). Fig.~\ref{fig:Raman2}(b1) is for free-space excitation and collection on the antenna.
A clear Fano lineshape is observed for the integrated counts of each peak as the laser is tuned over the hybrid resonance, consistent with the expectation that the SERS enhancement is the product of a pump-field enhancement factor (resonator property at the pump frequency) and LDOS enhancement at the Stokes-shifted frequency~\cite{Ye2012}. Here, the pump enhancement that is common to all three lines shows the hybrid Fano-lineshape, while the LDOS at the Stokes-shifted frequencies is unstructured, and is essentially in the featureless wing of the broad Lorentzian antenna resonance (as sketched in Fig.~\ref{fig:sketch}). The Fano lineshape in the pump enhancement occurs because  free-space pumping  directly drives the broad antenna resonance, and its coupling to the narrow resonance appears as a transparency line~\cite{Pan2020,Shlesinger2021}. Panel (c1) shows results on the same hybrid, but now driving directly the cavity through the waveguide, by aligning the pump to one of the gratings, but still collecting from the antenna. When we sweep the laser frequency, all three Raman lines again follow a very similar behavior, due to a strong spectral modification of the pump enhancement, yet an unstructured LDOS at the vibrational lines. Instead of a Fano lineshape, the SERS enhancement now traces an almost Lorentzian resonance which reports the transfer function for pump light from the grating coupler, via the cavity, to the antenna (see Fig.~S5(b) in the SM~\cite{Note1}).
We now turn to the second scenario, wherein the hybrid provides selective LDOS enhancement, instead of pump enhancement (right column of Fig.~\ref{fig:Raman2}). To this end, we select a different cavity with a longer  $\lambda_c=$\SI{800}{\nano\meter}. In contrast to the previous scenario, now the pump is not tuned through the hybrid resonance but through a wavelength interval blue-detuned by about 25 nm, exactly such that only the \SI{473}{\centi\meter}$^{-1}$ BPT Raman line is scanned across the hybrid resonance (same structure as in Fig.~\ref{fig:Raman1st}(c)).
The system is again first interrogated by free-space pumping and collection. In stark contrast to the earlier results, now only one Raman line shows a Fano lineshape, while the other Raman lines follow a broad shoulder.
Indeed, the pump field now enjoys the broadband enhancement of the antenna, without modification by the cavity mode, but the hybrid LDOS now shows a Fano line that acts only on the vibration that it is resonant with (see Fig.~\ref{fig:sketch}). The shoulder in the other lines is due to the typical SERS enhancement of the bare antenna, decreasing as one tunes away from resonance~\cite{McFarland2005}. 

Finally, Fig.~\ref{fig:Raman2}(c2) shows waveguide addressed SERS. As the photonic crystal cavity prohibits optical transport between the grating coupler and antenna except on cavity resonance, the pump is now provided from free-space, while the Raman signal is collected from the outcoupling grating. When the \SI{473}{\centi\meter}$^{-1}$ peak is tuned over the hybrid resonance it shows a strong Lorentzian response whereas the 1073 and 1576 cm$^{-1}$ lines barely pass through the waveguide. Here, the pump enhancement still originates from the broad antenna resonance, while the hybrid resonance provides both the LDOS and the collection efficiency into the waveguide required to enhance and efficiently collect only a single Raman line. 
Significant SERS enhancement into the cavity port concurs with the Fano dip in the free-space signal due to suppressed radiative loss.
Importantly, the detected SERS counts from integrated operation are as good as for free-space addressing which translates into an estimated 7-fold stronger integrated SERS enhancement than obtained for free-space addressing when taking into account the poor grating coupling efficiencies (see SM~\cite{Note1} for further analysis of signal levels).
Comparison with a semi-analytical model detailed in the SM~\cite{Note1}, predicts a hybrid cooperativity (comparing the cavity-antenna coupling rate with the optical losses) of order 1, ensuring a good trade-off between hybrid enhancement and collection rates~\cite{Doeleman2016Antenna-CavityLinewidth,Shlesinger2021}.

To conclude, we reported on the realization of a new generation of hybrid plasmonic-dielectric cavity resonators, with a lithographically designed gap antenna accurately coupled to a narrow linewidth photonic crystal cavity. The hybrid resonances allow one to selectively enhance and collect single Raman lines in a guided mode, paving the way for on-chip applications of spectrometer-free, specific Raman species detection.  
The intense resonances obtained with the 20 nm gap modes, and the narrow linewidth obtained through hybridization with the high-Q cavity mode are particularly interesting for optomechanical strong coupling~\cite{Dezfouli2019MolecularModes} and sideband-resolved molecular optomechanics. For instance, sideband resolution is highly relevant to observe parametric instabilities in the few molecule regime~\cite{Lombardi2018}. Achieving this regime would require further optimization of cavity fabrication to reach higher quality factors as expected for photonic crystal geometries~\cite{Quan2011}. Indeed, we estimate the optomechanical cooperativity $C_m$, which determines the degree of coherence of the light-vibration interaction, to be on the order of $10^{-3}$ for our system (see SM for details~\cite{Note1}). Cooperativities $C_m\gtrsim1$ are expected for challenging but realistically achievable parameters $Q_c\simeq5000$ and 4 nm dimer gaps.

\begin{acknowledgments}
We thank Alejandro Mart\'{\i}nez and Ewold Verhagen for stimulating discussions.  This work is part of the Research Program of the Netherlands Organization for Scientific Research (NWO). The authors acknowledge support from the European Unions Horizon 2020 research and innovation program under Grant Agreements No. 829067 (FET Open THOR).
\end{acknowledgments}

%

\end{document}


\title{Supplementary Information - Integrated Sideband-Resolved SERS with a Dimer on a Nanobeam Hybrid}

\author{Ilan Shlesinger}
\thanks{Equal contribution}
\affiliation{Department of Physics of Information in Matter and Center for Nanophotonics, NWO-I Institute AMOLF, Science Park 104, NL1098XG Amsterdam, The Netherlands}
\author{Isabelle M. Palstra}
\thanks{Equal contribution}
\affiliation{Institute of Physics, University of Amsterdam, 1098 XH Amsterdam, The Netherlands}
\affiliation{Department of Physics of Information in Matter and Center for Nanophotonics, NWO-I Institute AMOLF, Science Park 104, NL1098XG Amsterdam, The Netherlands}
\author{A. Femius Koenderink}
\email{f.koenderink@amolf.nl}
\affiliation{Department of Physics of Information in Matter and Center for Nanophotonics, NWO-I Institute AMOLF, Science Park 104, NL1098XG Amsterdam, The Netherlands}
\affiliation{Institute of Physics, University of Amsterdam, 1098 XH Amsterdam, The Netherlands}

\date{\today}

\maketitle
\newpage


\begin{figure}[b]
  \centering
  \includegraphics[width=0.7\textwidth]{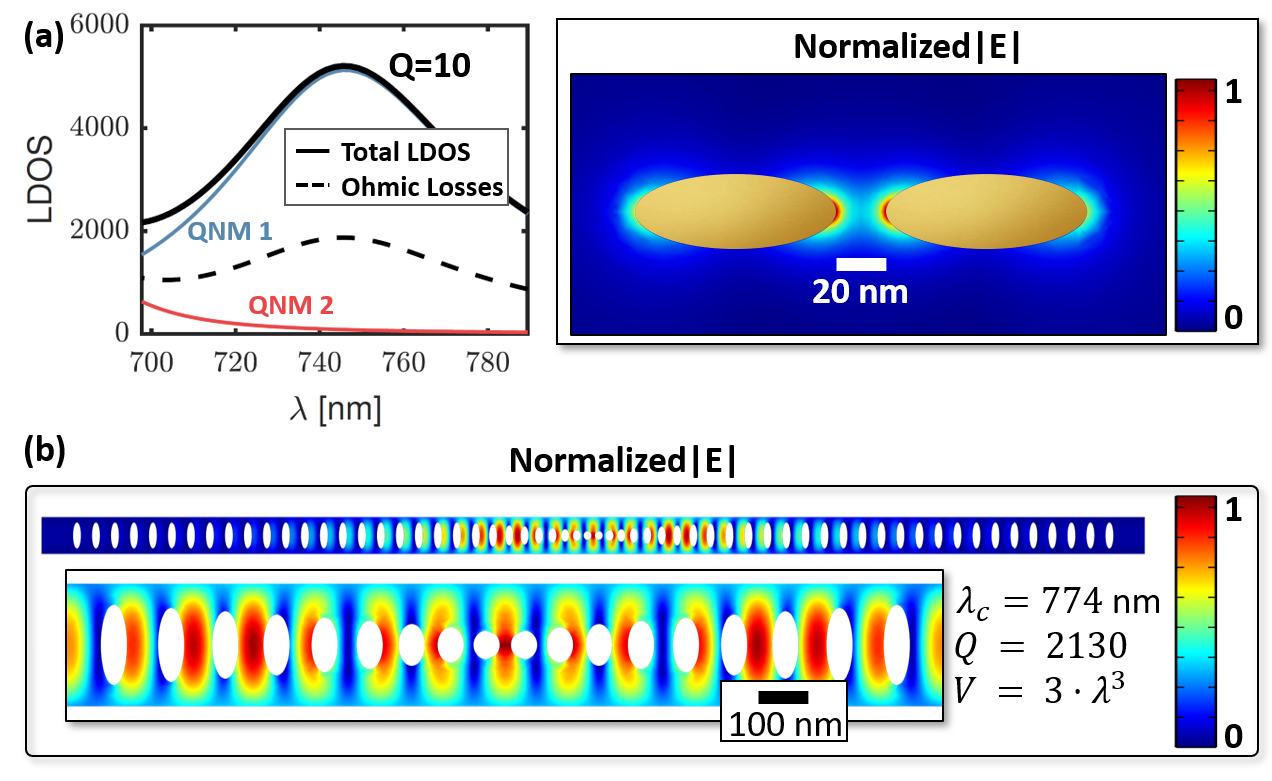}
  \caption{ FEM simulation of the dimer antenna (a) and the photonic crystal nanobeam (b). (a) From FEM simulations one can calculate the modes (lossy quasi-normal modes QNMs~\cite{Yan2018}), and deduce the LDOS for a dipole situated in the middle of the dimer gap. The dipolar mode (QNM 1) is the main contributor for the LDOS and we can extract a mode volume $|V|=10^{-4}\lambda^3$ and a quality factor of Q=10 matching the darkfield spectra. (b) An example of design of the photonic crystal nanobeam made of silicon nitride ($n_\mathrm{SiN}=4$) and surrounded by glass ($n_\mathrm{SiO_2}=1.47$), tuned to $\lambda_c=774\,\mathrm{nm}$ and resulting in a Q=2130 and a mode volume $V=3(\lambda_c/n_\mathrm{SiO_2})^3$.
  }
  \label{fig:sup7}
\end{figure} 

\begin{figure}[b]
  \centering
  \includegraphics[width=0.7\textwidth]{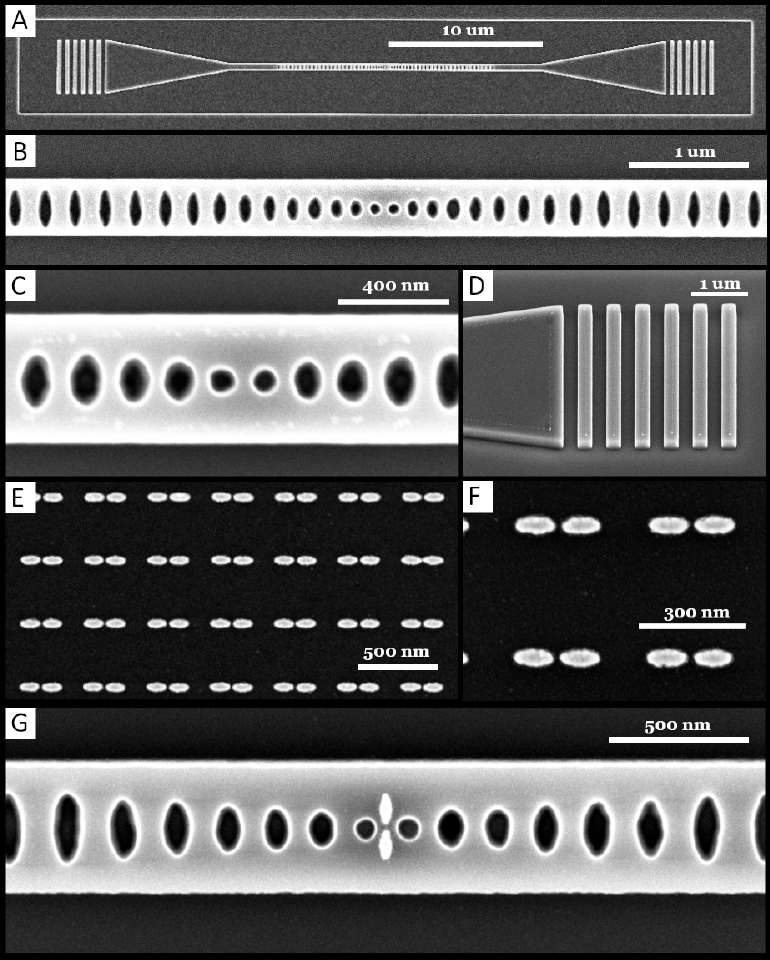}
  \caption{SEM images of the individual constituents, bare cavity (A-C), grating coupler (D), dimer antennas (E,F), and of the hybrid structure (G).
  }
  \label{fig:sup1b}
\end{figure}

\textbf{Antenna and Cavity design - } The antennas are dimers of gold ellipsoids with a short and long axis of 40 nm and 80 nm respectively, and 40 nm thick. Their long axes lie in one line, and they have a gap of around 20 nm. FEM simulations are shown in Fig.\ref{fig:sup7}. The gold is modelled with a two pole Lorentz-Drude permittivity. From an eigenfrequency simulation using the MAN package for COMSOL~\cite{Yan2018} we obtain the modes of the system, and extract a quality factor of Q=10 and a mode volume of $|V|=10^{-4}\lambda^3$.
The nanobeam cavity design consists of a 440 nm wide beam, which has two sets of 20 elliptical holes with long axis 153 nm, short axis 51 nm, and pitch 228 nm that form Bragg mirrors on either side of a cavity (Fig.\ref{fig:sup7}(b)). The cavity is formed by two additional sets of 10 holes, where the long axis and pitch taper down linearly to 51 nm and 143 nm, respectively, resulting in a cavity resonance within the photonic bandgap. An example of field profile for the cavity mode is shown in Fig.\ref{fig:sup7}(b).
To couple the light in and out of the system, we taper out the width of the waveguide to 2.5 µm, followed by a grating. The gratings serve for out-of-plane incoupling and outcoupling of light, and each consist of 6 grating lines of Si$_3$N$_4$ 250 nm wide, with a pitch of 500 nm (see Fig.~\ref{fig:sup1b}(D)).

\begin{figure}[b]
  \centering
  \includegraphics[width=0.8\textwidth]{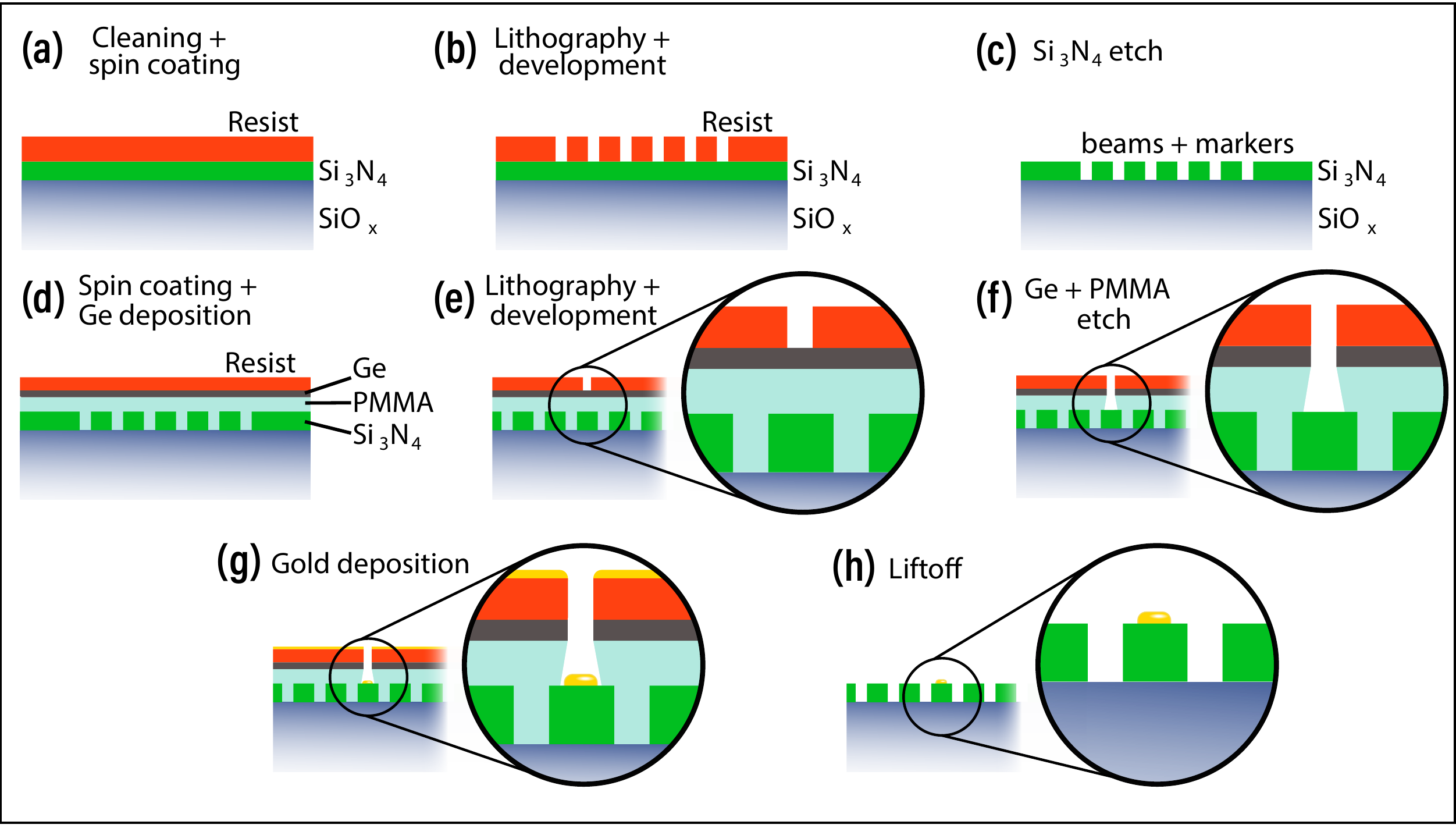}
  \caption{Fabrication steps of the dimer on nanobeam hybrid.
  }
  \label{fig:sup1}
\end{figure}

\textbf{Fabrication - } A two-step lithography process allows us to fabricate the dimer on top of the center of the photonic crystal cavity. The full procedure is sketched in Fig.~\ref{fig:sup1}.

(a-c) The cavity is fabricated by etching a photonic crystal into Si$_3$N$_4$ on a glass substrate. The  wafers (LioniX International) are made of a silicon support layer around 1.5 mm thick, with 8 µm of glass SiO2 topped with a 210 nm layer of Si$_3$N$_4$. (d-h) A subsequent lithography step using a Ge hard mask allows fabricating a dimer antenna, on top of the photonic crystal cavity. The antennas follow the design described above. Markers created in the first step allow for the precise alignment of the dimer and the cavity center for proper hybridization of the two optical modes.
SEM images of the bare cavity, the bare antennas, the grating couplers and the full hybrid structure are shown in Fig.~\ref{fig:sup1b}.

To functionalize the antennas with the Raman active molecules, the samples are immersed for 12 hours in 1 mM solution of
biphenyl-4-thiol (BPT, Sigma-Aldrich, 97\%) in anhydrous ethanol
(Sigma-Aldrich, $<$0.003\% H2O). The BPT molecules will form a self-assembled monolayer (SAM) on the surface of the gold~\cite{Vericat2014,Ahmed2021}. At the end of the incubation period, the samples are rinsed multiple times with ethanol and DI water to remove any unbound thiols and finally dried under N2 blow.
After functionalization, the system is encapsulated in PMMA to provide the host index required for the nanobeam mode. Raman spectrum of powder BPT shows minor differences with the spectrum obtained for SAMs on gold~\cite{Xomalis2020}.

\begin{figure}
  \centering
  \includegraphics[width=0.9\textwidth]{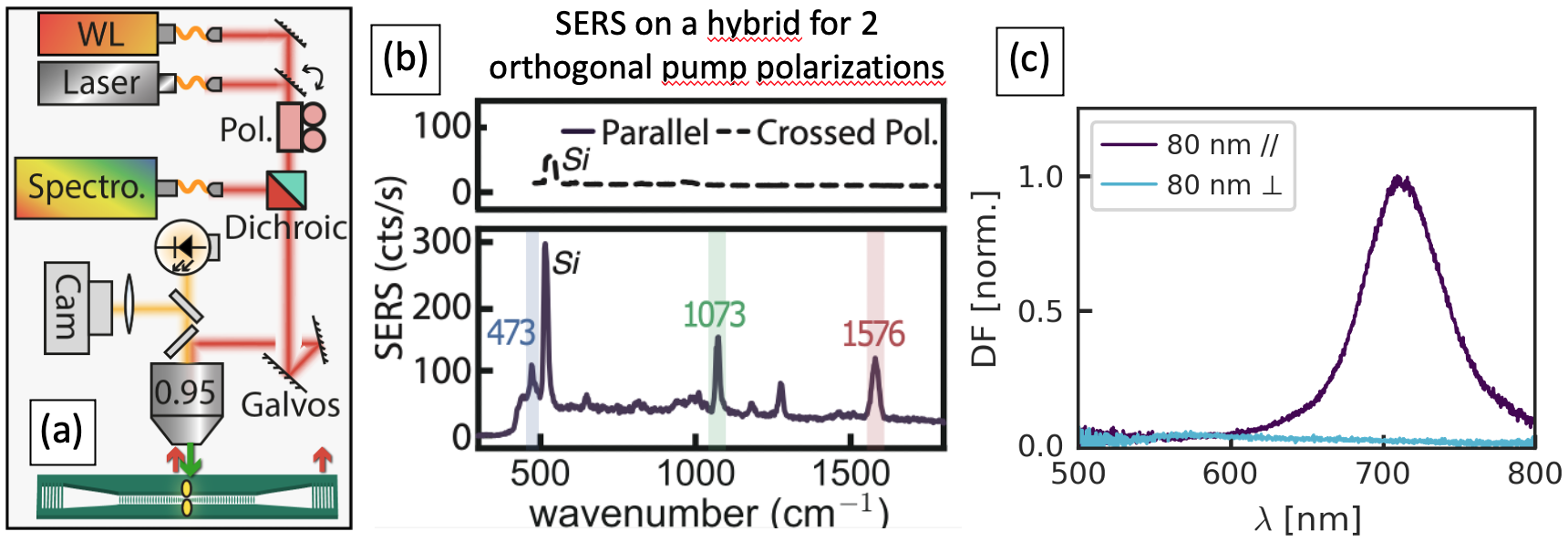}
  \caption{(a) Schematic of the home-built confocal optical Raman spectroscopy setup. (b) SERS spectrum for orthogonal (top) and parallel (bottom) polarization of the laser pump w.r.t. the dimer long axis. The reference case (orthogonal polarization) was obtained after 20 min integration time, resulting in a saturation of the CCD for the main Si Raman peak (520 $\mathrm{cm}^{-1}$).
  (c) Two DF spectra for incident light polarized parallel or orthogonal to the long axis of the dimers. The sample is a field of dimer antennas on glass at a distance of 2 um from each other, covered with PMMA. Measurements are performed with a Nikon LV-UEPI2 epi illumination unit, using a 100 W halogen light source and a Nikon 20X, NA0.45, DF objective. The collected light is injected into a 200 um optical fiber connected to a spectrometer (Avantes 2048TEC-2-USB2, fan-cooled CCD). Approximately 20 antennas lie on the collection spot. 
  }
  \label{fig:sup2}
\end{figure}

\textbf{Optical setup - }
The measurements are performed on a home-built confocal Raman spectroscopy setup~\ref{fig:sup2}(a). The signal is collected with a microscope objective (Olympus MPlan IR, 100X, NA= 0.95) and routed to a fiber-coupled spectrometer (Andor Shamrock A-SR-303i-B-SIL) equipped with a cooled silicon CCD camera (Andor iVac A-DR324B-FI). Excitation is performed with a narrowband tunable diode laser (Toptica DL-Pro 780), spectrally cleaned with a pair of bandpass filters (Semrock TBp01-790/12). Prior to insertion of detected light into the collection fiber, Rayleigh scattered light is filtered with a set of two notch filters (Thorlabs NF785-33). This configuration allows for the detection of both Stokes and anti-Stokes signals, where the pump frequency is tunable over tens of nanometers, covering the relevant range. 
We quantify the cavity resonances by waveguide transmission spectra, focusing light from a fiber-coupled white-light (WL) halogen lamp on one grating coupler and collecting from the other.

\textbf{Lower bound on Raman enhancement - }
A lower bound on the Raman enhancement of the hybrid can be obtained by comparing the counts obtained when the hybrid is excited with those of a reference case. The anisotropy of the dimer antenna results in a very different resonance frequency for the plasmonic mode depending on the polarization of the electric field, and only a laser beam polarized along the long axis of the dimers can excite the dimer resonance from the far field as shown in the DF spectra of Fig.~\ref{fig:sup2}(c).
A reference case for each hybrid resonator is then obtained by simply rotating the polarization of the excitation laser so as to be orthogonal to the long axis of the dimer. A measurement with a 20 minute integration time (limited by small mechanical drifts at larger time scales) is shown in Fig.~\ref{fig:sup2}(b). Only the Raman peaks from the silicon (the $520\, \mathrm{cm}^{-1}$ saturating the CCD) are visible, with no signal from the BPT (lower than the noise background of about 1ct/s). Compared to the BPT signal obtained from the same sample when the dimer is excited with a parallel polarization, we deduce a minimum SERS enhancement of the hybrid of the order of $1.1\times10^4$. Using a semi-analytical model, we expect this value to be much lower than the actual enhancement, which we predict to be on the order of $10^{6-7}$ (see section ``Simulation'' below). It is important to note that Raman signal is detected from the BPT molecules (and the electronic background from the gold) only when the laser is linearly polarized and centered over the antennas. Confocal scanning of the excitation collection thus results in a diffraction limited spot at the dimer position. No signal from the BPT, or even the PMMA layer is observed at any other position (except from the grating couplers in the crossed configuration, where the antenna is excited).

\begin{figure}
  \centering
  \includegraphics[width=\textwidth]{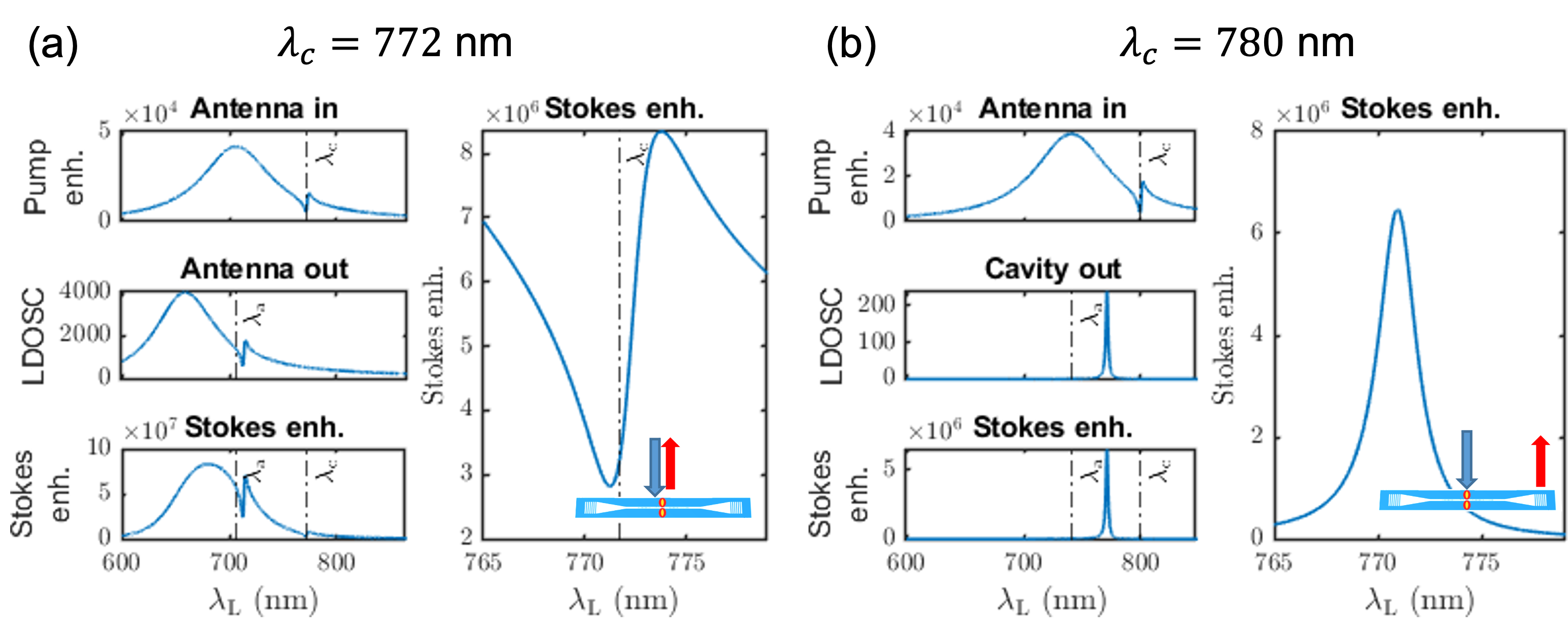}
  \caption{Stokes enhancement vs pump wavelength obtained from a semi-analytical model. (a) corresponds to the hybrid enhancing the pump for input and output through the antenna, and (b) to the hybrid enhancing the LDOS, with an output through the cavity grating. The total Stokes enhancement (lower left panel in each case) is a product of pump enhancement (top left panel), and collected LDOS enhancement in the corresponding port (middle left panel) calculated at a frequency shifted from the laser by a the molecule's vibrational frequency. The zoomed Stokes enhancement for both hybrids (right panel) correspond to the dashed line plots of Fig.~5 (b1, c2) in the main text.
  }
  \label{fig:sup3}
\end{figure}

\begin{figure}
  \centering
  \includegraphics[width=0.4\textwidth]{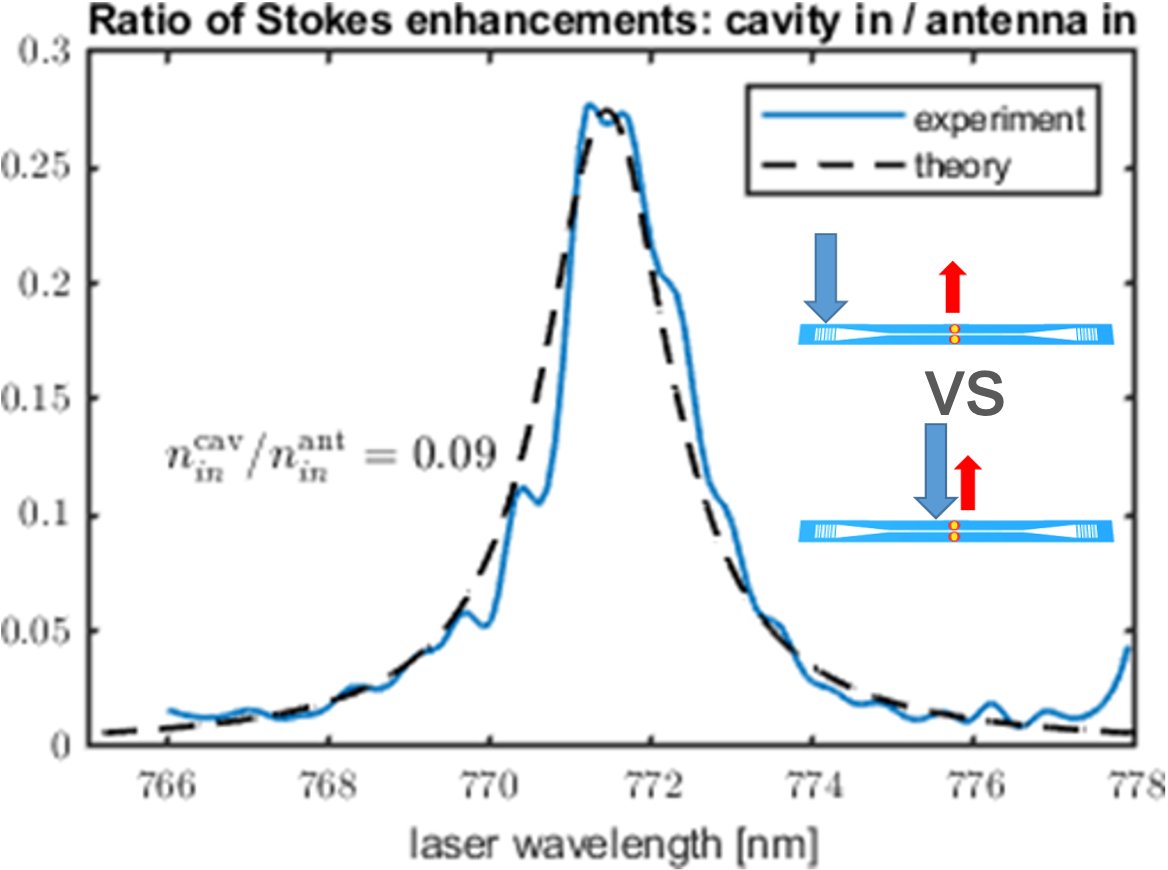}
  \caption{Determining the input and output coupling from SERS enhancement ratios in different configurations. The input coupling efficiency is determined by fitting the ratio of SERS enhancement for same output configuration but different inputs. 
  }
  \label{fig:sup4}
\end{figure}

\textbf{Simulation - }
The SERS enhancement of the hybrid with two optical modes and different input and output ports can be compared to a semi-analytical model presented elsewhere~\cite{Shlesinger2021}. This model is based on semi-classical Langevin equations describing multiple optical modes with different input and output ports, coupled to a mechanical resonance. The model is fitted to the experimental data, with initial values obtained from finite-element COMSOL simulations. Importantly, both direct (antenna in, antenna out) and crossed configurations (with input or output through the waveguide) are simultaneously fitted with the same parameters for each hybrid. The simulated Stokes enhancement for the two hybrids studied in the main text are shown in Fig.~\ref{fig:sup3} (for a single input-output port configuration for each hybrid as an example).
The Stokes enhancement predicted by the model is on the order of $10^{6-7}$ and is almost equal for a collection through the antenna or a collection in the waveguide.
Most of the parameters can be estimated from WL measurements and FEM calculations, and the final values are summarized in table~\ref{table:1}.
\begin{table}[ht]
\centering
\begin{tabular}{||c | c | c | c||} 
 \hline
 Parameters & Cavity & Antenna & Comment \\ [0.5ex] 
 \hline\hline
 Frequency [THz] & 389 & 425 & From WL and DF \\ 
 \hline
 Q & 380 & 10 & WL and DF, fit for cavity \\
 \hline
 Mode Volume (in $\lambda^3$) & 12 & $10^{-4}$ & COMSOL, fit for the cavity \\
 \hline
 Ohmic losses $\gamma_i$ [2$\pi\times \mathrm{THz}$] & - & 19.9 & COMSOL \\
 \hline
 Radiative losses $\gamma_\mathrm{rad}$ [2$\pi\times \mathrm{THz}$] & - & 23 & DF \\
  \hline
 Incoupling efficiency $\eta_\mathrm{in}$ & 0.022 & 0.24 & NA=0.95, see text \\  \hline
 Outcoupling efficiency $\eta_\mathrm{out}$ & 0.13 & 0.24 & idem \\ [1ex] 
 \hline
\end{tabular}
\caption{Simulation parameters for the beam with a cavity resonant to the pump. Similar values are obtained for the beam with a hybrid enhancement of the LDOS. The source of the parameter value is given in the last column. When fitted, the parameters are constrained to values close to initial values obtained from FEM simulation.}
\label{table:1}
\end{table}
This leaves the input and output coupling efficiencies of the antennas which are estimated from the NA of the measuring objective. For before we said 0.95 as used in this study, we get $\eta\simeq0.24$ for the antenna port. These are subsequently used to estimate the waveguide coupling efficiencies. For this, we compare the SERS ratio obtained for the same hybrid for identical output conditions but with an input from the grating or an input through the antenna, which depend directly on the input efficiency ratio $\eta_\mathrm{in}^\mathbf{cav}/\eta_\mathrm{in}^\mathbf{ant}$. By comparing to the theoretical model one can correctly fit this ratio with only the cavity input coupling efficiency as a free parameter (other parameters fixed from previous fit routines). Fig.~\ref{fig:sup4} shows the resulting plot allowing to extract $\eta_\mathrm{in}^\mathbf{cav}$, and we obtain $\eta_\mathrm{in}^\mathbf{cav}=0.09 \,\eta_\mathrm{in}^\mathbf{ant}$. The incoupling efficiency to the waveguide is very inefficient under the current experimental conditions, since the grating is excited with the same objective and thus a very focused spot (in the order of $\lambda$) far from the optimal plane wave case expected for gratings. The same procedure is applied to extract the output coupling efficiency and we obtain a value of $\eta_\mathrm{out}^\mathbf{cav}=0.52 \,  \eta_\mathrm{out}^\mathbf{ant}$, reflecting the fact that the collection of light scattered from the grating is more effective than the injection of light.

\begin{figure}
  \centering
  \includegraphics[width=0.6\textwidth]{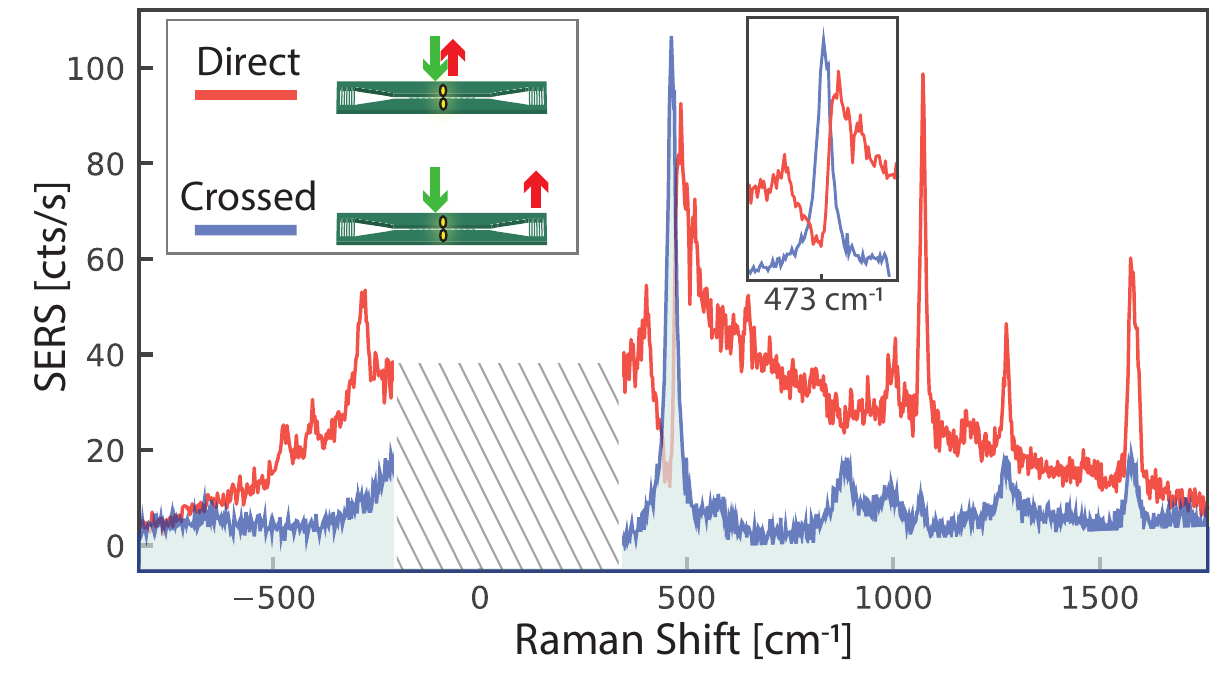}
  \caption{Integrated SERS vs free-space. Single shot SERS for the LDOS enhancement hybrid when pumping the antenna (400 uW pump power) and for antenna output (direct configuration) and waveguide output (crossed configuration).
  }
  \label{fig:sup5}
\end{figure}

\textbf{Integrated SERS efficiency - } In Fig.~\ref{fig:sup5} the Raman spectrum enhanced by the same hybrid is shown for both free-space and integrated configuration.
A first observation is that integrated SERS shows no background from the silicon nitride~\cite{Peyskens2018}, which is a unique feature of the hybrid resonator. Indeed, as opposed to simple geometries with antennas placed directly on a waveguide, here the cavity works as a filter for the pump, and only the peak resonant to the hybridized cavity mode (strong peak at $473\,\mathrm{cm^{-1}}$) is transmitted. Second, the good hybrid coupling allows for good emission through the cavity into the waveguide, so that an equivalent SERS enhancement is obtained for both free-space and integrated configuration. The best overall coupling is, in fact, obtained for hybrid cooperativities $4J^2/\kappa\gamma_a$ on the order of 1~\cite{Shlesinger2021}, where $J\propto1/\sqrt{V_c}$ is the cavity-plasmon coupling rate and and $\kappa$ and $\gamma$ the total losses of the cavity and antenna, respectively. This is due to a trade-off between improving the exchange rate in the cavity mode when using higher Q cavities, and the lower outcoupling rate due to smaller losses~\cite{Doeleman2016Antenna-CavityLinewidth,Shlesinger2021}. With the parameters obtained from the simulation, we estimate the cooperativity of these hybrids to be around $C\simeq 1.6$, ideally suited to enhance integrated operation.

\begin{figure}
  \centering
  \includegraphics[width=0.97\textwidth]{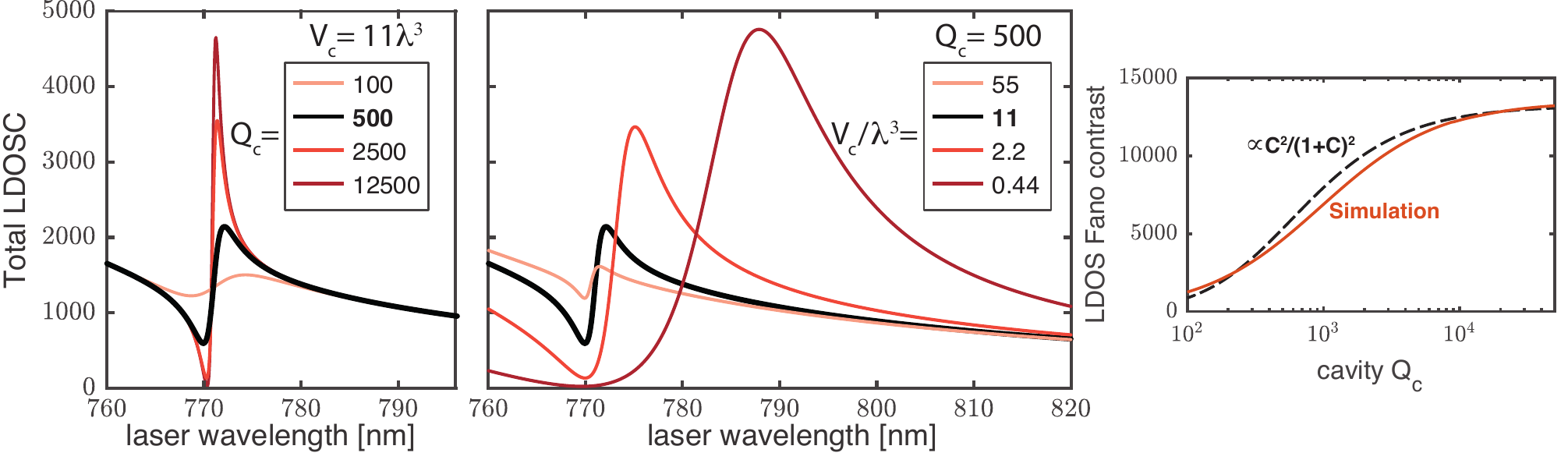}
  \caption{ Total (free-space and waveguide) collected LDOS vs cavity parameters: quality factor on the left and mode volume in the middle. The Fano contrast is mainly determined by the hybrid cooperativity, i.e. by the ratio $Q_c/V_c$. Right panel shows how the LDOS can be estimated by a scaling of $C^2/(C+1)^2$ of the maximum contrast ($C\rightarrow \infty$).
  }
  \label{fig:sup6}
\end{figure}

It is worth mentioning that the outcoupling from the WG is here mainly limited by the grating design, which wasn't optimized for this study. Furthermore, to take full advantage of the integrated behavior of the waveguide port, the signal should be collected by an optical fiber and not converted back into free-space as performed here for experimental simplicity. Coupling efficiencies above 80\% from chip to fiber can be obtained when using tapered coupled fibers as has been shown by other groups~\cite{Daveau17}. With this output collection efficiency for the guided signal, the counts obtained in the integrated configuration would be around 7 times higher than what is shown in Fig.~\ref{fig:sup6}, thus largely beating the free-space (direct) configuration.

\textbf{LDOS dependence on cavity parameters - } The total LDOS of the cavity is also directly related to the hybrid cooperativity, with a Fano contrast scaling as $C^2/(C+1)^2$. Fig.~\ref{fig:sup6} shows how the total collected LDOS depends on both the cavity quality factor and
mode volume. It can be seen that the maximum LDOS is dependent only on the ratio $Q_c/V_c$, which is proportional to $J^2/\kappa$ and thus to cooperativity. To reach higher enhancement values while retaining small linewidths it is still preferable to increase $Q_c$ rather than reducing $V_c$.

\textbf{Towards coherent molecular optomechanics - } The hybrid resonator enables sideband-resolved operation while maintaining good field enhancements. This paves the way towards the observation of new phenomena in SERS with few molecules, which was previously only restricted to cavity optomechanical systems. These include cooling, parametric amplification or quantum state control. Besides sideband resolution, the system needs to exhibit a good optomechanical cooperativity. For a single optical cavity mode, it is written~\cite{Bowen2015}:
\begin{equation}
    C_m=4\frac{n_\mathrm{cav} g_{0,m}^2}{\kappa\Gamma_m}\,,
\end{equation}
with $n_\mathrm{cav}$ and $\kappa$ the optical cavity's number of photons and linewidth, $\Gamma_m$ the mechanical linewidth, and $g_{0,m}$ the optomechanical coupling rate. The latter can be expressed for SERS systems as a function of the Raman activity and the mode volume of the optical mode~\cite{Roelli2016MolecularScattering,Schmidt2016QuantumCavities}. Here we estimate the optomechanical cooperativity by considering the hybridized cavity mode as the only optical resonance, with modified linewidth and mode volume $\kappa_\mathrm{H}=\kappa_c+2J^2 \mathrm{Im}(\chi_a(\omega_c))$ and $V_\mathrm{H}=V_c/(2|1+J\chi_a\sqrt{V_c/V_a}|^2)$, with $V_a$ the antenna mode volume at the molecule's position and $\chi_a$ the antenna susceptibility~\cite{Doeleman2016Antenna-CavityLinewidth,Shlesinger2021}. 
Using these values, we estimate the optomechanical cooperativity in our system to be $C_m\simeq 10^{-3}$. We have considered a laser power of 1 mW (deciding on the cavity photon number, here of 92 photons, and restricted to values below damage threshold as observed experimentally), for 1000 molecules (increasing by $\sqrt{N}$ the optomechanical coupling~\cite{Zhang2020}),  Raman activity of $500\si{\angstrom}^4/\mathrm{amu^{-1}}$~\cite{Roelli2016MolecularScattering} and typical optomechanical quality factor $Q_m=200$~\cite{LeRu2009PrinciplesSpectroscopy}.

Although the cooperativity with this particular system is still too low to observe nonlinear effects, by only increasing $Q_c$ to 5000 and with a realistic gap of 4 nm for the dimer, we expect $C_m\simeq 1.6$, large enough to allow for coherent optomechanical interaction and quantum control. These constraints could be further relaxed by using pulsed excitation~\cite{Lombardi2018}.

%